**Elastic properties of freely suspended MoS$_2$ nanosheets**

*Andres Castellanos-Gomez*[1,2,*], *Menno Poot*[1], *Gary A. Steele*[1], *Herre S.J. van der Zant*[1], *Nicolás Agraït*[2,3] and *Gabino Rubio-Bollinger*[2,*].

[1] Kavli Institute of Nanoscience, Delft University of Technology, Lorentzweg 1, 2628 CJ Delft (The Netherlands).
[2] Departamento de Física de la Materia Condensada. Universidad Autónoma de Madrid, Campus de Cantoblanco. E-28049 Madrid (Spain).
[3] Instituto Madrileño de Estudios Avanzados en Nanociencia IMDEA-Nanociencia. E-28049 Madrid (Spain).

E-mail: a.castellanosgomez@tudelft.nl , gabino.rubio@uam.es

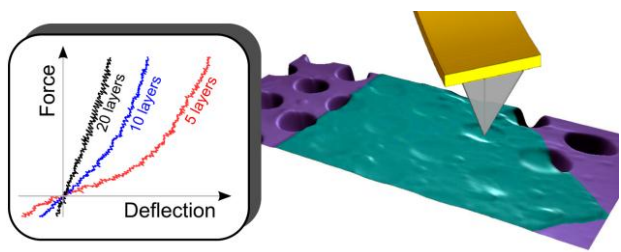

*We study the elastic deformation of few layers (5 to 25) thick freely suspended MoS$_2$ nanosheets by means of a nanoscopic version of a bending test experiment, carried out with the tip of an atomic force microscope. The Young's modulus of these nanosheets is extremely high (E = 0.33 TPa), comparable to that of graphene oxide, and the deflections are reversible up to tens of nanometers.*

Two-dimensional crystals are promising materials for next-generation flexible electronic devices. Indeed graphene, which exhibits a very high mobility, has been recently applied as transparent and flexible electrode.[1, 2] The lack of a bandgap in pristine graphene, however, hampers its application in semiconducting devices. Up to now, two different strategies have been employed to fabricate semiconducting two-dimensional crystals. While the first one relies on opening a bandgap in graphene through top-down engineering[3, 4] or chemical modification,[5] the second one involves the use of another two-dimensional crystal with a large intrinsic bandgap.[6, 7] Atomically thin crystals of the semiconducting transition metal dichalcogenide molybdenum disulphide (MoS$_2$) have emerged as a very interesting substitute/complement to graphene in semiconducting applications due to its large intrinsic bandgap of 1.8 eV [8-12] and high mobility μ > 200 cm$^2$V$^{-1}$s$^{-1}$.[13] Nevertheless, the mechanical properties of this nanomaterial, which will dictate their applicability in flexible electronic applications, remain unexplored so far.

In this context, we present measurement on the elastic properties of freely suspended MoS$_2$ nanosheets, with thicknesses ranging from 5 to 25 layers, in a bending test experiment performed with the tip of an atomic force microscope (AFM). These measurements allow us to determine simultaneously the Young's modulus (*E*) and the initial pre-tension (*T*) of these MoS$_2$ nanosheets.

**Figure1**(a) shows a contact mode AFM topography of a 5-7 layer thick MoS$_2$ flake deposited onto a 285 nm SiO$_2$/Si pre-patterned substrate. We have used the contact mode

AFM to determine the thickness and to characterize the topography of the deposited flakes. Our high resolution AFM measurements provide lattice resolution even in the suspended region of the MoS$_2$ flakes (see supplementary information) which demonstrates the very clean nature of our fabrication technique (see materials and methods section).

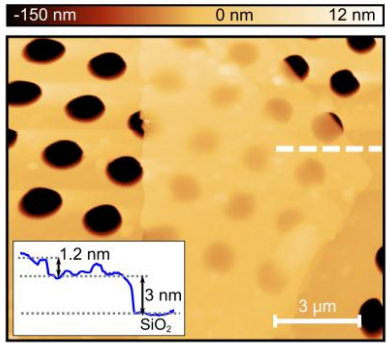

**Figure 1.** (a) Contact mode AFM topography of a 3 – 4.2 nm thick (5-7 layers) MoS$_2$ flake deposited on top of a 285 nm SiO$_2$/Si substrate pre-patterned with an array of holes 1.1 μm in diameter. (Inset) topographic line profile acquired along the dashed line.

Once the suspended nanosheet under study is characterized, we measure its elastic mechanical properties by using the tip of an AFM to apply a load cycle in the center of the suspended region of the nanosheet while its deflection is measured, as shown in **Figure 2**(a) When the tip and sample are in contact, the elastic deformation of the nanosheet ($\delta$), the deflection of the AFM cantilever ($\Delta z_c$) and the displacement of the scanning piezotube of the AFM ($\Delta z_{piezo}$) is related by

$$\delta = \Delta z_{piezo} - \Delta z_c \qquad (1)$$

The force applied is related to the cantilever deflection as $F = k_c \cdot \Delta z_c$, where $k_c$ is the spring constant of the cantilever ($k_c = 0.88 \pm 0.20$ N/m [14]).

Figure 2(b) presents typical force vs. deflection traces ($F(\delta)$ traces hereafter) measured at the center of the suspended part of MoS$_2$ nanosheets. The shape of the traces is clearly thickness dependent: the thinnest sheets (5-8 layers) present strongly nonlinear $F(\delta)$ traces, while sheets thicker than 10 layers $F(\delta)$ traces are linear. This can be explained when considering that the mechanics of these suspended nanolayers results from the tradeoff between plate (bending-dominated) and membrane (stretching-dominated) behavior. In a simplified continuum mechanics model of the sheet, the relationship between the applied force at the center of the flake and the resulting deformation of the suspended nanosheet is [15, 16]

$$F = \left[\frac{4\pi E}{3(1-v^2)} \cdot \left(\frac{t^3}{R^2}\right)\right]\delta + (\pi T)\delta + \left(\frac{q^3 E t}{R^2}\right)\delta^3 \qquad (2)$$

where $E$ and $v$ are the Young's modulus and the Poisson ratio ($v = 0.125$),[17] $t$ is the thickness of the nanosheet, $R$ is its radius, $T$ its pre-tension and $q = 1/(1.05 - 0.15v - 0.16v^2)$. The first term in expression (2) corresponds to the mechanical behavior of a plate with a certain bending rigidity.[15, 16] The second term represents the mechanical behavior of a stretched membrane.[18] Finally, the third term takes into account the stiffening of the layer during the force load cycle which makes $F(\delta)$ nonlinear.[15] The cubic thickness dependence ($t^3$) of the bending rigidity makes it the most relevant term for the thicker sheets. On the contrary, the static and deflection-induced tension terms are the most significant ones for ultrathin sheets, explaining the observed crossover from non-linear to linear $F(\delta)$.

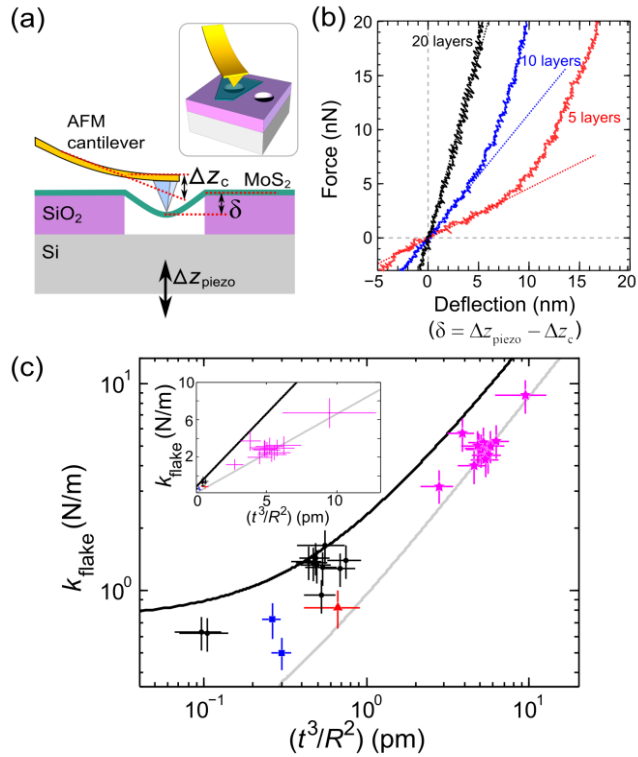

**Figure 2.** (a) Schematic diagram of the nanoscopic bending test experiment carried out on a freely suspended MoS$_2$ nanosheet. (b) Force vs. deflection traces measured at the center of the suspended part of MoS$_2$ nanosheets with 5, 10 and 20 layers in thickness. The slope of the traces around zero deflection is marked by a dotted line. (c) Elastic constant vs. $t^3R^{-2}$ measured for 26 MoS$_2$ suspended nanosheets with thickness ranging from 25 down to 5 layers. Data points sharing color and symbol correspond to suspended nanosheets of the same MoS$_2$ flake. The relationship $k_{\text{flake}}$ vs. $t^3R^{-2}$ calculated with expression (3) for $E$ = 0.21 TPa and $T$ = 0.03 N/m (gray solid line) and $E$ = 0.37 TPa and $T$ = 0.23 N/m (black solid line). The insert in (c) shows the same graph on a linear scale.

In the limit of small deformation, the relationship between the force and the deflection of the flake is linear and its spring constant is

$$k_{\text{flake}} = \left.\frac{\partial F}{\partial \delta}\right|_{\delta=0} = \frac{4\pi E}{3(1-\nu^2)}\cdot\left(\frac{t^3}{R^2}\right) + \pi T, \qquad (3)$$

which scales as $t^3R^{-2}$. Figure 2(c) shows $k_{\text{flake}}$ vs. $t^3R^{-2}$ for 26 MoS$_2$ suspended nanosheets with thicknesses from 25 down to 5 layers. Using Eq. (3) we find that all measurements are bounded between $E$ = 0.21-0.37 TPa and $T$ = 0.03-0.23 N/m .

For non-linear $F(\delta)$ (i.e. thin flakes), one can extract both the Young's modulus ($E$) and the initial pretension ($T$) of a suspended MoS$_2$ sheet independently by fitting expression (2) to it. **Figure 3**(a) shows a $F(\delta)$ for a 8 layer nanosheet from which we obtain $E$ = 0.35 ± 0.03 TPa and $T$ = 0.05 ± 0.02 N/m. Note that the deflections of the suspended sheets are reversible up to several tens of nanometers.

For the thicker flakes $F(\delta)$ is linear, but one can obtain both the Young's modulus ($E$) and the initial pretension ($T$) using the method as described in Ref. [19]. Figure 4(b) shows the radially averaged compliance ($1/k_{flake}$) of the 8 layers thick $MoS_2$ nanosheet also measured in Figure 4(a). The best fit yields $E = 0.40 \pm 0.03$ TPa and the pre-tension $T = 0.02 \pm 0.02$ N/m, which are in agreement with the ones obtained from the non-linear force *vs*. deflection traces.

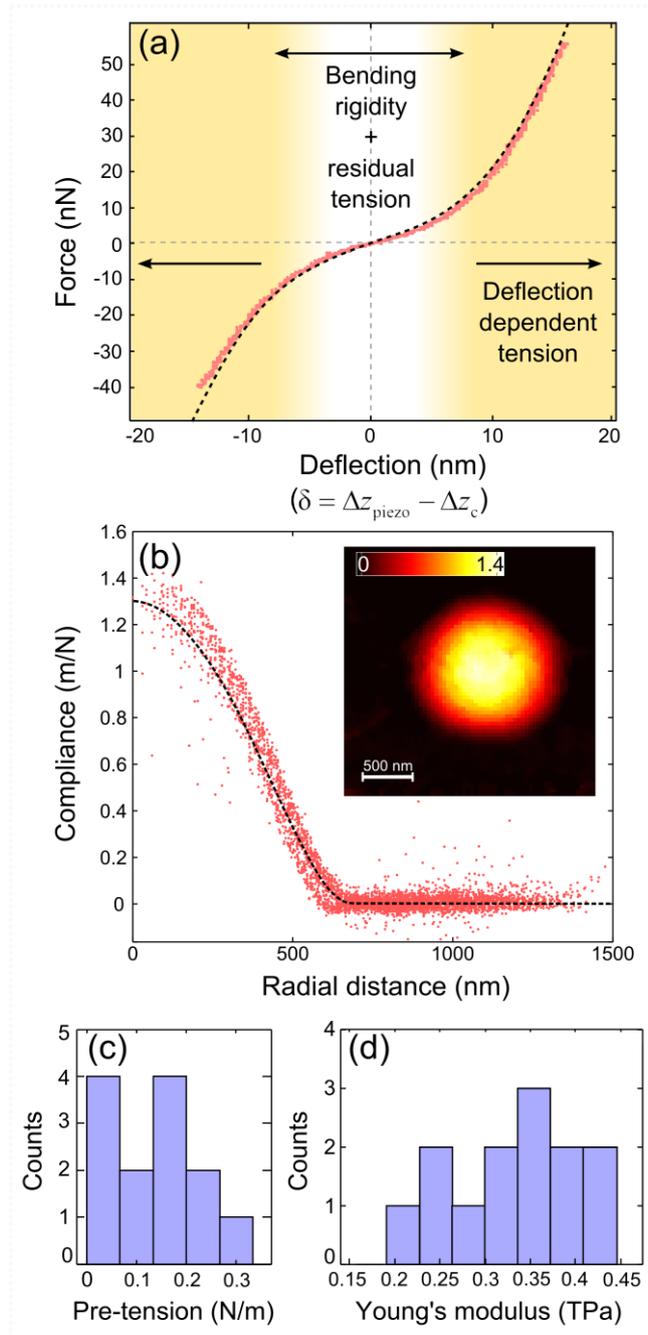

**Figure 3.** (a) Force *vs*. deflection traces obtained on a flake 8 layers thick suspended over a hole 1.1μm in diameter. The dotted black trace is the fit to expression (2), employed to obtain the Young modulus $E = 0.35 \pm 0.02$ TPa and the pre-tension $T = 0.05 \pm 0.02$ N/m of this nanolayer. (b) Force-volume measurement showing a colormap of the compliance (inset) and its radially-averaged profile of the sheet as (a). The solid line shows a fit using the model from Ref. 18. (b) Histogram of the initial pre-tension obtained from the fit to expression (2) for 13 sheets 5 to 10 layers thick. (c) Histogram of the Young's modulus obtained from fitting $F(d)$ curves to expression (2) for the same 13 sheets plotted in panel (b)).

The described bending test experiment allows resolving flake-to-flake variation in $E$ and $T$. These are attributed to different density of defects in the flakes and/or adhesion force with the substrate. To determine the dispersion of the mechanical properties of our suspended sheets we have carried out a statistical analysis of several suspended MoS$_2$ sheets. Figure 3(b) and (c) show the histogram of the pre-tension and Young's modulus values obtained for 13 membranes 5 to 10 layers thick by fitting the force *vs*. deflection traces to expression (2).

The mean Young's modulus and pre-tension and their standard deviation are $E = 0.33 \pm 0.07$ TPa and T = $0.13 \pm 0.10$ N/m . This Young's modulus value is extremely high, only one third lower than exfoliated graphene (one of the stiffest materials on Earth with $E = 0.8 - 1.0$ TPa)[18, 20] and higher than other 2D crystals such as graphene oxide (0.2 TPa),[21, 22] hexagonal boron nitride (0.25 TPa)[23], carbon nanosheets (10 – 50 GPa)[24] or some 2D clays (22 GPa).[25] It is also remarkable that the variation in *E* is restrained between 0.21 TPa and 0.42 TPa, indicating a high homogeneity of the MoS$_2$ flakes. The variation is much smaller than the one observed for other 2D crystals such as graphene (0.02 – 3 TPa),[19] graphene oxide (0.08 – 0.7 TPa)[21] and Na$_{0.5}$ - fluorohectorite (10 – 30 GPa).[25]

The Young's modulus we obtained for ultrathin MoS$_2$ flakes ($E = 0.33 \pm 0.07$ TPa) should be compared with the value for bulk MoS$_2$, $E_{bulk} = 0.24$ TPa[26] . This raises again the controversial question whether the Young's modulus is size dependent or not.[27] For one-dimensional systems, the measured discrepancies with respect to the bulk value are attributed to surface effects.[27] In the case of layered materials like the MoS$_2$, this discrepancy is attributed to the presence of stacking faults which are dominant in the mechanical properties of these materials. Therefore, the thinner the nanosheet the lower the presence of stacking faults, allowing the study of the intrinsic mechanical properties of the material. In our case, a low density of stacking faults would explain the high Young's modulus observed, indicating that the thickness of our MoS$_2$ nanolayers is lower than the average distance between stacking faults.

In summary, we have fabricated and characterized freely suspended MoS$_2$ nanosheets 5 to 25 layers thick and we have studied their mechanical properties. A continuum mechanics model has been used to account for the experimentally observed mechanical behavior of suspended MoS$_2$ sheets with thickness down to just 5 layers. The average Young's modulus of these suspended nanosheets is extraordinary high $E = 0.33 \pm 0.07$ TPa, comparable to that of graphene oxide. We have also found that these suspended nanolayers present low pre-strain which is rather uniform across the studied flakes with different thicknesses. Further, these suspended sheets are very tough: they can stand deformations up to tens of nanometers elastically without breaking. The low pre-tension and high elasticity and Young's modulus of these crystals makes them attractive substitutes or alternatives for graphene in applications requiring flexible semiconductor materials.

*Experimental*

*MoS$_2$ flakes fabrication*

Although scotch-tape-based micromechanical cleavage can be used to fabricate atomically thin MoS$_2$ crystals [28], the traces of adhesive glue left during this procedure can alter the properties of the fabricated crystals and may eventually contaminate the microscope tip used in AFM measurements [29]. Therefore, it is preferable to fabricate clean samples using an all-dry procedure based on poly (dimethyl)-siloxane (PDMS) stamps [30-32], commonly used in soft-lithography [33, 34], instead of adhesive tape. In order to fabricate freely suspended atomically thin MoS$_2$ flakes we first cleave a bulk MoS$_2$ crystal by pressing the PDMS stamp surface against the crystal and subsequently we rapidly peel off the stamp. The cleaved flakes are then transferred to a pre-patterned oxidized silicon wafer [35] with circular holes 1.1 μm in diameter and 200 nm deep by pressing the stamp against the SiO$_2$ surface and slowly peeling it off (~5 s to peel off the stamp completely from the surface), favoring the transfer of the flakes from the PDMS stamp to the SiO$_2$ surface. As the flakes are transferred onto a pre-patterned substrate with holes, further lithography steps are not necessary to fabricate the suspended nanolayers. In this way we avoid contamination due to the exposure to lithographic resists [36, 37].

*MoS$_2$ optical identification*

MoS$_2$ flakes can be readily identified using an optical microscope. In particular, flakes less than 30 nm thick can be easily located because their optical contrast under red illumination ($\lambda = 600 \pm 5$ nm) decreases almost linearly with the number of layers [31]. We used a *Nikon Eclipse LV100* optical microscope under normal incidence with a 50× objective (numerical aperture NA = 0.8) and with a digital camera EO-1918C 1/1.8" (from *Edmund Optics*) attached to the microscope trinocular. The illumination wavelength was selected by means of narrow band-pass filters (10 nm full with half maximum) purchased from *Edmund Optics*.

*AFM imaging*

After locating the flakes on the surface, a *Nanotec Cervantes* AFM (*Nanotec Electronica*) operated at room temperature and pressure has been used to study the topography of the selected nanosheets. Contact mode AFM has been chosen instead of dynamic AFM modes to avoid possible artefacts in the flake thickness measurements [38].

*Acknowledgements*

A.C-G. acknowledges fellowship support from the Comunidad de Madrid (Spain). This work was supported by MICINN (Spain) through the programs MAT2008-01735, MAT2011-25046 and CONSOLIDER-INGENIO-2010 'Nanociencia Molecular' CSD-2007-00010, Comunidad de Madrid through program Nanobiomagnet S2009/MAT-1726 and the European Union (RODIN, FP7).

*Re*ferences

# Supporting information:

# Elastic properties of freely suspended MoS$_2$ nanosheets

*Andres Castellanos-Gomez*[1,2],*, *Menno Poot*[1], *Gary A. Steele*[1], *Herre S.J. van der Zant*[1], *Nicolás Agraït*[2,3] and *Gabino Rubio-Bollinger*[2],*.

[1] Kavli Institute of Nanoscience, Delft University of Technology, Lorentzweg 1, 2628 CJ Delft (The Netherlands).
[2] Departamento de Física de la Materia Condensada. Universidad Autónoma de Madrid, Campus de Cantoblanco. E-28049 Madrid (Spain).
[3] Instituto Madrileño de Estudios Avanzados en Nanociencia IMDEA-Nanociencia. E-28049 Madrid (Spain).

E-mail: a.castellanosgomez@tudelft.nl , gabino.rubio@uam.es

**Sample preparation procedure:**

The PDMS stamps have been fabricated using the silicone elastomer kit Sylgard 184 (from Dow Corning) (see supplementary information in Ref. [1]). Very briefly:

(1) The 184 Sylgard polymer base and the curing agent are mixed in a 10:1 ratio by weight.

(2) A Teflon mold is pressed against a clean and flat surface like a glass slide or a silicon chip. Note that the part of the stamp we will use to cleave the crystals is the one in contact with this flat and clean surface (working surface).

(3) The mixed 184 Sylgard base and curing agent is poured into the mold.

(4) A vacuum desiccator is used to degas the mixed 184 Sylgard. In this way we remove air bubbles and the surface of the stamp will be more flat. Fifth, the filled mold is placed in an oven at 60 ºC during 24 hours.

(5) The stamp is unmolded. A Teflon mold is used to facilitate the unmolding process without breaking the stamp.

**Sample characterization:**

*Optical microscopy:*

To optically identify few-layer thick MoS$_2$ flakes we have used a *Nikon Eclipse LV100* microscope under normal incidence with a 50x objective (0.8 numeric aperture). The optical micrographs were acquired with a *Canon EOS 550D* digital camera attached to the optical microscope. Flakes with regions with different thicknesses present different apparent color as explained in ref. [1]. Figure S1 shows an optical image of a multilayered MoS$_2$ crystal deposited on top of a pre-patterned substrate.

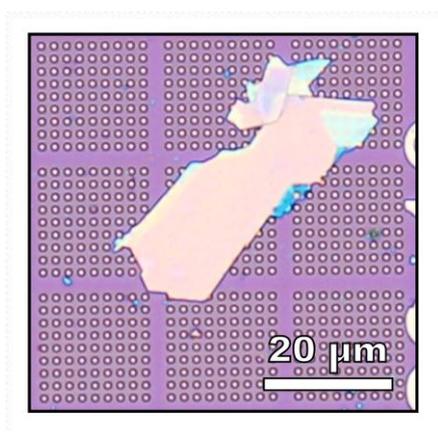

**Figure S1**. Optical micrograph of a MoS$_2$ flake deposited on a pre-patterned substrate with holes. The flake present regions with different apparent colors which correspond to different thicknesses as checked by atomic force microscopy (AFM). Note that there are regions very transparent in which it is possible to identify the holes underneath. These regions are typically thinner than 10 layers.

*Atomic force microscopy:*

The AFM characterization of the flakes and the bending test experiments have been carried out with a *Cervantes* AFM from *Nanotec Electrónica* operated at room pressure and temperature conditions. The thickness of the flakes has been determined by contact mode AFM to avoid possible artifacts in the flake thickness measurements.[2]

We have additionally found that high resolution AFM measurements provide lattice resolution even in the suspended region of the MoS$_2$ flakes which demonstrates the very clean nature of our fabrication technique (see materials and methods section). Figure 1(b) shows a normal force map, measured in constant height AFM mode, obtained in the suspended part of a 3 nm thick (5 layers) MoS$_2$ flake. In this figure one can resolve the sulphur atoms from the surface spaced 0.3 nm, in agreement with previous STM measurements in bulk crystals.[3]

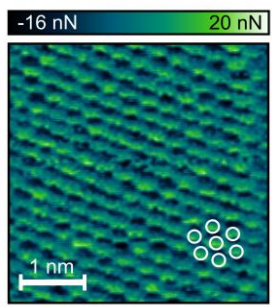

**Figure S2.** Normal force map measured in constant height AFM mode at the center of the suspended region of a 3 nm thick MoS$_2$ flake. In this image one can resolve the spacing (0.3 nm) between sulfur atoms at the surface.

**Bending test experiment:**

*Force versus displacement traces:*

To obtain the force vs. deflection traces (Figure 2(b) and Figure 4(a)) in the manuscript) we first measure force vs. displacement curves and we employ expression (1) from the original manuscript. To determine the AFM cantilever deflection $\Delta z_c$, the AFM photodetector is calibrated measuring force *vs.* displacement traces on the SiO$_2$/Si substrate. Once the AFM tip is in contact with the hard substrate, a displacement of the AFM scanning piezo $\Delta z_{piezo}$ (which itself is calibrated by measuring the height of monoatomic steps of graphite) produces a deflection of the cantilever ($\Delta z_c$) equal to $\Delta z_{piezo}$.

Figure S3 shows some typical force vs. displacement traces measured on the SiO$_2$ substrate (panel a) and at the center of suspended sheets with of 5, 10 and 20 layers in thickness.

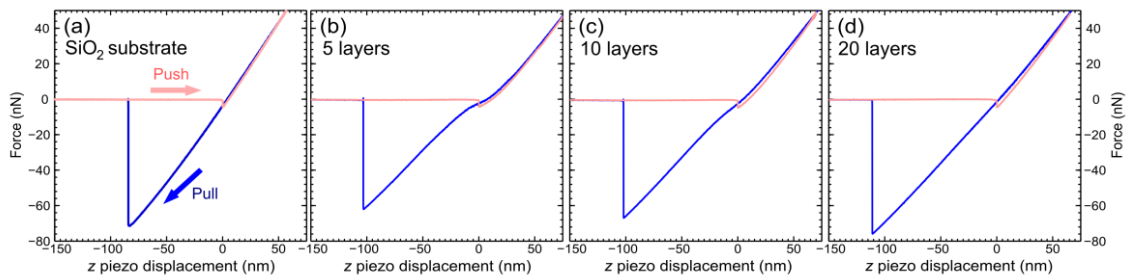

**Figure S3**. (a) to (d) are force *vs.* displacement traces measured on the SiO$_2$ substrate (a) and at the center of the suspended part of MoS$_2$ nanosheets with 5 layers (b), 10 layers (c) and 20 layers (d) in thickness.

*Radially averaged compliance maps: force-volume method*

For the thicker flakes that show only linear $F(\delta)$ one can obtain both the Young's modulus ($E$) and the initial pretension ($T$) in an alternative way from a compliance image of the suspended layers, such as shown in the inset in Figure 3(b) in the original manuscript.[4] Following the procedure described in Ref. [4], the deflection of a circular membrane with Young's Modulus $E$, pre-tension $T$ and radius $R$ was calculated for a point load $F$ applied at position ($r_0,\theta_0$). This gives the induced deflection profile $\delta(r,\theta; r_0,\theta_0)$. The local compliance is then given by the ratio of the deflection and the applied force at that point: $\delta(r,\theta; r_0,\theta_0)/F$. For a circular hole the compliance is independent of $\theta_0$ and its radial profile depends $F$ only on three parameters: the hole radius $R$, the bending rigidity (related to the Young's modulus, thickness and Poisson's ratio), and the tension $T$. Least squares is used to fit the calculated compliance profiles $1/k_{flake}(r_0)$ to the experimental ones to independently determine the Young's modulus and pre-tension of a few-layer suspended MoS$_2$.

This alternative procedure to obtain the Young's modulus and the pretension is relatively slow: the compliance maps (inset in Figure 3(b) in the manuscript) consist of 64 × 64 force *vs.* distance traces in the force-volume AFM mode. The advantage of this method is that one can determine independently the Young's modulus and the pre-tension of relatively thick suspended crystals even though their $F(\delta)$ are linear. On the other hand, the method that based on fitting the non-linear $F(\delta)$ traces to equation (2) of the manuscript (Figure 3(a) in the manuscript), is fast and works well for atomically thin (less than 10 layers) suspended MoS$_2$ crystals which are the ones more interesting for flexible semiconducting applications.

*Force dependent contact mode AFM topography:*

We have also used the AFM to study the deformation of the suspended layers when the AFM tip is applying a force during the scan. Figure S4 shows the AFM contact mode topography of a MoS$_2$ sheet suspended over a hole, measured with different set-point forces. When the set-point is positive the tip is pressed against the layer and the suspended membrane is deflected downwards. Also when a negative set-point is used, the suspended layer can be deflected upwards. Using this data one can reproduce, within experimental uncertainty, the force vs. deflection traces measured in the same nanolayer. The main drawback of this method relies on the fact that the zero force value drifts slowly with the time because of differential thermal expansion of several mechanical parts of the microscope. If this effect can be reduced/counteracted, this method can be employed to obtain compliance maps without using the force-volume mode (increasing considerably the speed of the measurement).

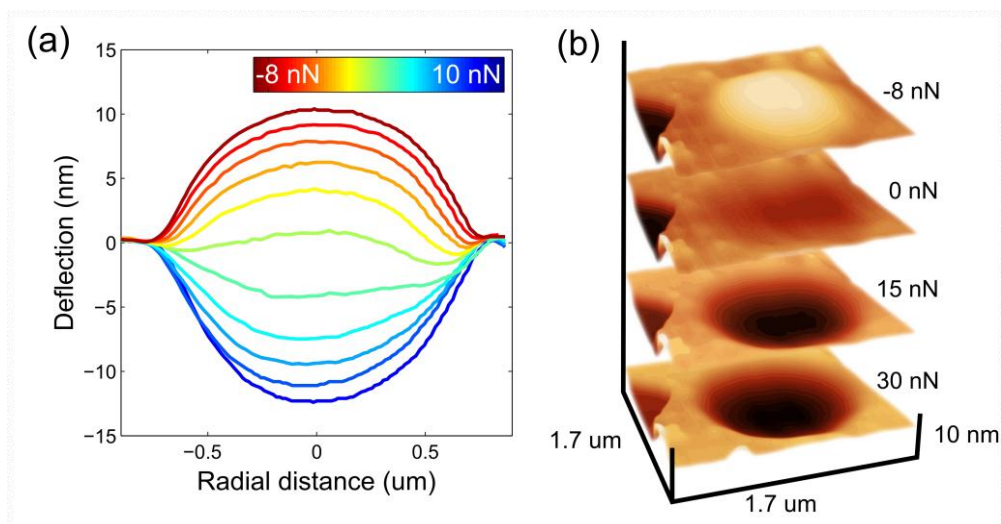

**Figure S4**. (a) AFM contact mode topographic line profiles measured along a suspended MoS$_2$ sheet (8 layer thick) with different force set-points (b) Three-dimensional representation of AFM topographies of the same suspended MoS$_2$ sheet acquired in contact mode at different set-point forces.